\documentclass[aps,pra,twocolumn,superscriptaddress,notitlepage]{revtex4-1}
\usepackage[colorlinks,linkcolor={blue},citecolor={blue},urlcolor={blue}]{hyperref} 
\usepackage{graphicx,amsfonts,amssymb,amsmath,hyperref,hypcap,enumerate}
\usepackage{subfigure}
\usepackage{amsmath}
\usepackage{braket}
\usepackage{bm}
\usepackage[T1]{fontenc}
\usepackage[utf8]{inputenc}
\usepackage{float}
\usepackage{amssymb} 
\usepackage{amsfonts}
\usepackage{xcolor}
\usepackage{tabularx}
\usepackage{multirow}
\usepackage{commath}
\usepackage{lineno}
\usepackage{soul}
\usepackage{scalerel}
\usepackage{adjustbox,stmaryrd}
\usepackage{array}
\usepackage{enumitem} 

\newcommand{\bs}{\boldsymbol}
\newcommand{\be}{\begin{equation}}
\newcommand{\ee}{\end{equation}}
\newcommand{\bmat}{\begin{pmatrix}}
\newcommand{\emat}{\end{pmatrix}}

\begin{document}
\title{Layer Hall counterflow as a model probe of magic-angle twisted bilayer graphene}

\author{Jihang Zhu}
\affiliation{Condensed Matter Theory Center and Joint Quantum Institute, Department of Physics, University of Maryland,
College Park, Maryland 20742, USA}
\affiliation{Max Planck Institute for the Physics of Complex Systems, 01187 Dresden, Germany}
\author{Dawei Zhai}
\affiliation{Department of Physics, The University of Hong Kong, Hong Kong, China}
\affiliation{HKU-UCAS Joint Institute of Theoretical and Computational Physics at Hong Kong, Hong Kong, China}
\author{Cong Xiao}
\affiliation{Institute of Applied Physics and Materials Engineering, University of Macau, Taipa, Macau SAR, China}
\author{Wang Yao}
\affiliation{Department of Physics, The University of Hong Kong, Hong Kong, China}
\affiliation{HKU-UCAS Joint Institute of Theoretical and Computational Physics at Hong Kong, Hong Kong, China}

\begin{abstract}
The recent constructions of flat moir\'e minibands in specifically twisted multilayer graphene and twisted transition metal dichalcogenides (TMDs) have facilitated the observation of strong correlations with a convenient tunability. 
These correlations in flat bands result in the band dispersion heavily influenced by carrier densities, leading to filling-dependent quasiparticle band renormalizations. Particularly, in magic-angle twisted bilayer graphene (MATBG), the band structure--including the quasiparticle energy and wavefunction--is crucial in understanding the correlated properties. Previous theoretical studies have demonstrated the presence of a time-reversal-even charge Hall counterflow in response to a direct current (DC) electric field in twisted bilayers as chiral structures.
In this study, we show that such layer Hall counterflow can serve as a sensitive probe for MATBG model parameters, which are currently ambiguous as a result of unavoidable structural relaxation and twist-angle disorder. 
We present the layer Hall counterflow and the associated in-plane magnetization for three different MATBG continuum models, based on which many-body interacting models have been widely applied to study strong correlations in MATBG.
At the single-particle level, our findings indicate notable differences in layer-projected Hall conductivity, both in magnitude and sign, between different MATBG continuum models. Furthermore, our self-consistent Hartree calculations, performed on each of these single-particle continuum models, reveal renormalized layer-projected Hall conductivity by the self-consistent Hartree field.
\end{abstract}

{\let\newpage\relax\maketitle}

\section{Introduction}
Recent advances in the construction of vertically stacked van der Waals (vdW) heterostructures have illuminated the peculiar properties of these materials when twisted. A central attraction of this research frontier is the complex moir\'e superlattice structure, an outcome of such a twist. A notable manifestation of this is the appearance of flat electronic bands in the moir\'e superlattices of, for example, graphene~\cite{BM,TBGPRL2007,TBGMagicAngleOriginPRL2019} and TMDs~\cite{WuMacDonaldPRL2019,HongyiNSR2020,ZhaiPRM2020,TwistedWSe2MagicFuLiang2021}. These flat bands provide a platform for various interaction-driven quantum phenomena~\cite{moireReviewEvaMacDonaldNatMater2020,moireReviewNatPhysBalents2020,moireReviewRubioNatPhys2021,moireReviewExptFolksNatRevMat2021,moireReviewJeanieLauNature2022,moireexcitonreviewNature2021,moireexcitonreviewNatRevMater2022}.
Moreover, the structural chirality inherent in twisted vdW heterostructures has enabled the exploration of various chiral effects. 
This is epitomized by the observation of strong circular dichroism (CD) in chirally twisted graphene stacks~\cite{CDTBGNatNano2016}, attributed to the emergence of an in-plane magnetic dipole moment $\bs{m}_{\parallel}$
in the chiral stacks accompanying the longitudinal current induced by the electric field of light.

The observation of CD in chiral graphene stacks has stimulated many theoretical studies on the chiral optical responses of twisted bilayers~\cite{Brey2DM2017,TobiasPRL2018,TobiasPRB2018,CDTBGslidingMelePRB2019,TobiasPRB2020,CDTwistedhBNPRL2020,OpticalTBGVietnamPRR2020,TobiasNanoscale2020,TobiasPRB2021,OpticalDopedTBGPRB2022,Optical2DwithTRSShandongPRB2023,OpticalTBGVietnamPRB2023}.
The optical conductivity tensor $\sigma(\omega)$
for a chiral bilayer has been established using a modified Kubo formula~\cite{Brey2DM2017}. A pivotal discovery in this context has been the identification of the off-diagonal component, $\sigma_{xy}(\omega)$, which is responsible for the emergence of CD.
It describes a Hall-drag-like process, i.e., the transverse current response in one of the layers induced by the electric field in the other layer.
As explicitly formulated in Ref.~\cite{TobiasPRL2018}, the transverse currents dictated by $\sigma_{xy}(\omega)$ in the two layers flow in opposite directions, offering an intuitive understanding of the emergence of a longitudinal in-plane magnetic dipole moment, $\bs{m}_{\parallel}=d_0\hat{\bs{z}}\times(\bs{j}_{\perp}^{1}-\bs{j}_{\perp}^{2})/2$ ($d_0$ is the interlayer distance, and the superscript denotes the layer).

The optical responses of a chiral bilayer have also been formulated in terms of electromagnetic coupling~\cite{TobiasPRL2018,TobiasPRB2018,TobiasPRB2021}, which extended the study of optical properties of twisted bilayer graphene (TBG) beyond CD.
Chiral plasmon modes characterized by a $\bs{m}_{\parallel}$ accompanying the longitudinal current have been proposed~\cite{TobiasPRL2018}.
Furthermore, an in-plane magnetic field can provoke a $\bs{m}_{\parallel}$ via the counterflow conductivity $\sigma_{\text{cf}}(\omega)$~\cite{BM,TobiasPRL2018,TobiasPRB2018}, or a longitudinal current via the chiral conductivity $\sigma_{xy}(\omega)$~\cite{TobiasPRL2018,TobiasPRB2018,TobiasPRB2020}. These responses have been studied at the zero-frequency limit, characterized by the Drude weights $D_{i}=\text{lim}_{\omega\rightarrow0}\,\omega\,\text{Im}\{\sigma_{i}(\omega)\}$, and the chiral Drude weight $D_{xy}$ in the context of TBG was first evaluated for large twist angles~\cite{TobiasPRL2018,TobiasPRB2018} and around the magic angle~\cite{TobiasPRB2020}.

A finite chiral Drude weight implies the presence of opposite DC electric Hall transport in the two layers. It is noted that the emergence of a linear Hall current under time-reversal (TR) symmetry in each individual layer, being {\it non-isolated}, shall not be viewed as a violation of the Onsager's reciprocity relation~\cite{DZhai_LayerHall_2023}.
The existence of linear Hall counterflow in a TR invariant bilayer has been substantiated by its prediction with significant magnitudes over a broad range of twist angles in TBG and twisted homobilayer TMDs~\cite{DZhai_LayerHall_2023}. The effect is rooted in a band geometric quantity: the $k$-space layer current vorticity $\bs{w}_{n}^{l}(\bs{k})\propto\partial_{\bs{k}}\times\bs{v}_{n}^{l}(\bs{k})$, where $n$ and $l$ are the band and layer index, respectively, 
and $\bs{v}_{n}^{l}$ is the band velocity projected to layer $l$ (see Sec.~\ref{sec_LayerHallCond}).
This geometric quantity is a characteristic of chiral structures and its weight below the Fermi energy $E_F$ yields the Hall conductivity in layer $l$, i.e., $\sigma_{\text{H}}^{l}\propto \tau\sum_{\varepsilon_{n\bs{k}}<E_{F}} w_{n}^{l}(\bs{k})$, where $\tau$ is the relaxation time.
The geometric origin of this Hall conductivity is reminiscent of the $k$-space Berry curvature contribution to the anomalous Hall conductivity~\cite{NiuRMP2010}. 

We note that the chiral Drude weight has been given in terms of a Fermi surface expression involving the layer current in Ref.~\cite{TobiasPRB2020}: $D_{xy}=(2A)^{-1}\sum_{n,\bs{k}}\hat{\bs{z}}\cdot(\bs{j}_{n\bs{k}}^{1}\times\bs{j}_{n\bs{k}}^{2})\delta(\varepsilon_{n\bs{k}}-E_{F})$, 
to characterize the electronic chirality. The exemplary effect discussed therein is the longitudinal current response to an adiabatically applied in-plane magnetic field in TBG.
Via integration by parts, one identifies that it is equivalent to $\sigma_{\text{H}}^{l}$ up to a factor that is the relaxation time~\cite{DZhai_LayerHall_2023}.
It is interesting to notice that different electromagnetic responses in chiral bilayers can be characterized by the same response coefficient, which is not unusual within the linear response framework.

So far, most of the existing literature on chiral responses in twisted bilayers has predominantly utilized single-particle approximations. For TBG, the Bistritzer-MacDonald (BM) model~\cite{BM} is a commonly employed theoretical framework. 
However, it is crucial to emphasize the complexity in MATBG modeling, as subtle variances even at the single-particle level can substantially influence the system's properties.
Specifically,
the presence of sample-dependent strain and twist-angle disorder introduces significant uncertainties in effective low-energy Hamiltonian parameters.
Such practical details have been shown to affect the electronic structures of MATBG dramatically, yet their influence on chiral responses has not been considered.
Another critical aspect is the dominant role of Coulomb interactions in the flat bands of MATBG, which lead to significant band renormalizations. Even though the effect of interactions has been taken into account in the longitudinal optical conductivity in TBG~\cite{OpticalTBGHartreePRB2020,OpticalTBGInteractionNPJQuantMater2020}, its impact on chiral responses remains unexplored.

In this work, we take into account all detailed structural relaxations and higher-order interlayer tunnelings, as well as self-consistent Hartree (SCH) fields, in the MATBG continuum model. We perform a systematic study on the layer Hall counterflow and its associated in-plane magnetization $\bs{m}_{\parallel}$, using three different MATBG continuum models: (i) the BM model~\cite{BM}, (ii) the generalized BM model with a non-local interlayer tunneling~\cite{Xie_WeakField_2021}, (iii) the comprehensive continuum model taking into account atomic relaxations using two sets of tight-binding model parameters~\cite{JKang_TBGmodel_2023, Vafek_TBGmodel_2023}. These MATBG continuum models are commonly used as the starting point for studying various many-body phenomena. 
We find that different terms in the MATBG continuum model can result in distinct or even opposite effects on the layer-contrasted Hall conductivity, among which the most important feature is associated with the band dispersion and band velocity near the moir\'e $\gamma$ point that is highly sensitive to the model details.
Moreover, the SCH potentials introduce changes in several features of the layer-contrasted Hall conductivity, the extent of which depends on the specific continuum model used.
Our results provide valuable insights into the understanding of different continuum models of MATBG from the perspective of chiral responses, and the correspondingly induced large $\bs{m}_{\parallel}$ might provide insights into the occurrence of current-induced magnetization switching in MATBG~\cite{MagnetizationSwitching1,MagnetizationSwitching2}.

This paper is organized as follows: Section~\ref{sec_LayerHallCond} provides an overview of the layer-contrasted Hall conductivity.
Section~\ref{sec_MATBGmodels} describes the specifics of the three MATBG continuum models, based on which the layer Hall conductivities are calculated and presented in 
Sec.~\ref{sec_para}.
In Sec.~\ref{sec_sch}, we show that the layer Hall conductivity is significantly renormalized by the filling-dependent SCH potentials, and the renormalization details are model-dependent.

\section{The layer Hall conductivity}\label{sec_LayerHallCond}
\begin{figure}[t]
\centering
\includegraphics[width=0.35\textwidth]{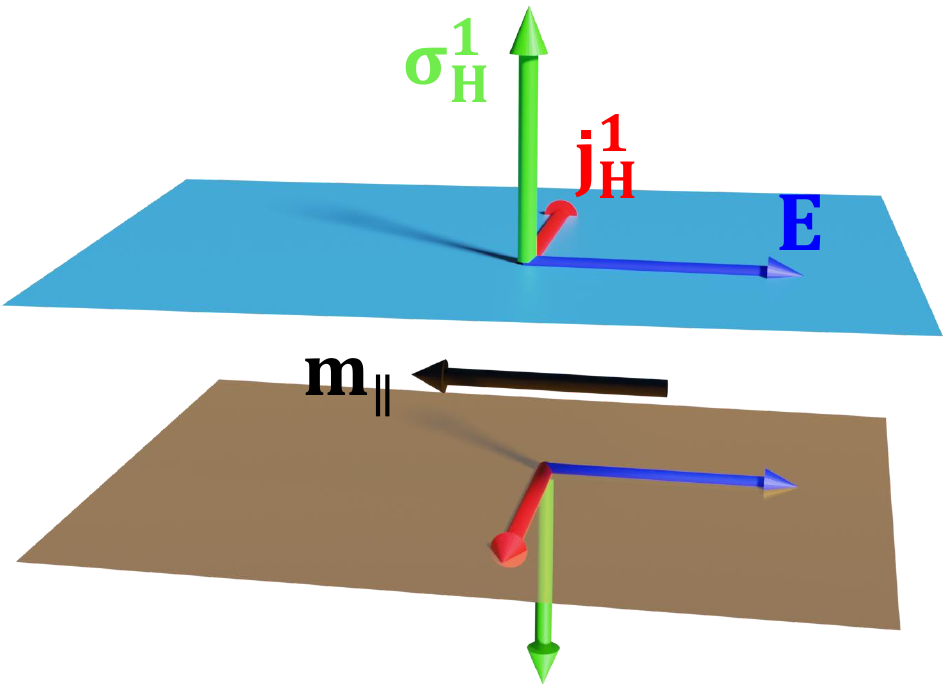}
\caption{\label{fig_m} {
  The chiral bilayer acquires an in-plane orbital magnetic moment $\bs{m}_{\parallel}$ (black) as a result of linear Hall counterflow (red) under an in-plane electric field (blue).
  }}
\end{figure}

In accordance with the semiclassical theory, when employing the relaxation-time approximation, the Hall conductivity that is projected onto layer $l$ of a nonmagnetic (TR symmetry preserved) chiral bilayer can be expressed as~\cite{DZhai_LayerHall_2023}
\begin{equation}\label{Eq_sigmaH_sea}
\sigma_{\rm H}^{l} 
= \frac{\tau e^2}{\hbar} \sum\limits_n \int \frac{d^2 \pmb{k}}{(2 \pi)^2}f_{0} \omega^l_n(\pmb{k}),
\end{equation}
where $\tau$ is the constant relaxation time, $f_{0} \equiv f_{0}(\varepsilon_{n\pmb{k}})$ is the equilibrium Fermi-Dirac distribution function and
\begin{equation}
\begin{split}
\omega_n^l(\pmb{k}) 
&= \frac{1}{2} \big[ \frac{\partial }{\partial \pmb{k}} \times \pmb{v}^{l}_n(\pmb{k}) \big]_z \\
&= \hbar {\rm Re} \sum\limits_{n' \neq n} \frac{\big[\pmb{v}_{nn'}(\pmb{k}) \times \pmb{v}^{l}_{n'n}(\pmb{k}) \big]_z}{\varepsilon_{n \pmb{k}} - \varepsilon_{n' \pmb{k}}}
\end{split}
\end{equation}
describes the momentum-space vorticity of the layer current,
$\pmb{v}^{l}_n(\pmb{k})=\braket{u_{n\bs{k}}|\frac{1}{2}\{P_l,\,\hat{\bs{v}}\}|u_{n\bs{k}}}$ is the projected band velocity of a Bloch state $\ket{u_{n\bs{k}}}$ onto layer $l$ with $P_1=\rm{diag}(1,\,0)$ and $P_2=\rm{diag}(0,\,1)$. The numerator of the second line of $\omega_n^l(\pmb{k})$ involves interband quantities, highlighting that non-zero $\sigma_{\rm H}^{l} $ requires layer hybridization. 
Integrated by parts, Eq.~(\ref{Eq_sigmaH_sea}) can be converted into a Fermi-surface form~\cite{DZhai_LayerHall_2023,TobiasPRB2020},
\begin{equation}\label{Eq_sigmaH_surface}
\sigma_{\rm H}^{l} 
= - \frac{\tau e^2}{2} \sum\limits_n \int \frac{d^2 \pmb{k}}{(2 \pi)^2} \frac{\partial f_{0}}{\partial \varepsilon_{n \pmb{k}}}
\big[ \pmb{v}_n(\pmb{k}) \times \pmb{v}^{l}_n(\pmb{k}) \big]_z,
\end{equation}
which involves the cross product of velocities in different layers.
In this form it is explicit that the effect can only stem from nonequilibrium kinetics of electrons around the Fermi surface in nonmagnetic systems, and the resulting conductivity is even under TR.
Notably, the conductivity is nonzero only in a chiral bilayer, satisfying 
$\sigma_{\rm H}^{1}=-\sigma_{\rm H}^{2}\ne0$, thus there are Hall currents flowing oppositely in the two layers: $\bs{j}_{\rm H}^{1}=-\bs{j}_{\rm H}^{2}=\sigma_{\rm H}^{1}\hat{\bs{z}}\times\bs{E}$. As the net Hall current vanishes, one may consider layer-resolved measurements by taking, for example, layer $1$ as the probed layer to extract the Hall current~\cite{DZhai_LayerHall_2023}.
Such Hall counterflow also contribute to a longitudinal in-plane magnetic dipole moment, $\bs{m}_{\parallel}=d_0\hat{\bs{z}}\times(\bs{j}_{\rm H}^{1}-\bs{j}_{\rm H}^{2})/2=-d_0\sigma_{\rm H}^{1}\bs{E}$, where $d_0$ is the interlayer distance.
These quantities are schematically depicted in Fig.~\ref{fig_m}. In the following calculations, we will show that the layer Hall conductivity $\sigma_{\rm H}^{l}$ could be very large in MATBG, leading to sizeable $\bs{m}_{\parallel}$ despite the interlayer distance is small.

\section{MATBG continuum models}\label{sec_MATBGmodels}
The BM model \cite{BM}, established over a decade ago, has been recognized for its practicality and accuracy, particularly in predicting the magic-angle at which low-energy bands become notably flat and strong correlations become dominant. Subsequent STM and transport experiments have led to several refinements to the BM model. Key modifications include: (i) the impact of out-of-plane relaxation \cite{TBGmodel_Koshino_2018, TBGmodel_Lucignano_2019, TBGmodel_Uchida_2014, TBGmodel_Wijk_2015, TBGmodel_Jain_2017, TBGmodel_Dai_2016}, separating the low-energy flat bands from remote ones;  (ii) the influence of in-plane relaxation \cite{ShiangFang_TBG_2019, TBGmodel_Koshino_2020, StephenCarr_TBG_2019, Vafek_TBGmodel_2023, JKang_TBGmodel_2023}, which substantially deforms the flat bands, and (iii) the higher-order gradient terms, which give rise to the particle-hole asymmetry \cite{ShiangFang_TBG_2019, Vafek_TBGmodel_2023, JKang_TBGmodel_2023}. 
Collectively, these effects are crucial for band structure renormalizations, stabilization of many-body ground states, and could be instrumental in elucidating and forecasting exotic phase transitions at fractional fillings. 

In the following subsections, we provide an overview of three widely-accepted MATBG continuum models: the original BM model \cite{BM} (including corrugation effects), the generalized BM model with the non-local interlayer tunneling \cite{Xie_WeakField_2021}, specifically the first-order gradient term, and a comprehensive continuum model taking into account all relaxation up to the second-order gradient terms \cite{JKang_TBGmodel_2023, Vafek_TBGmodel_2023}. Notably, this latter model integrates two sets of effective parameters, which are derived from earlier microscopic tight-binding models.

\subsection{The BM model}
\label{sec_MATBGmodels_BM}
The BM model Hamiltonian~\cite{BM}, widely used as a fundamental starting point for small-angle (< $10^\circ$) TBG calculations, of valley $K$ is given by
\begin{equation}
\label{Eq_H_BM}
\mathcal{H}_K = 
\begin{pmatrix}
h_D^{\theta/2}(\pmb{k} - \pmb{K}_1) & T(\pmb{r}) \\
T^\dagger(\pmb{r}) & h_D^{-\theta/2}(\pmb{k}' - \pmb{K}_2)
\end{pmatrix}.
\end{equation}
$h_D^{\theta}(\pmb{q})$ is the Dirac Hamiltonian twisted by $\theta$,
\begin{equation}
\begin{split}
h_D^{\theta}(\pmb{q}) 
&\approx (q_x + \theta q_y) \sigma_x + (q_y - \theta q_x) \sigma_y.
\end{split}
\end{equation}
$\pmb{K}_1$ and $\pmb{K}_2$ are Dirac points of first and second layers respectively in valley $K$.
The interlayer tunneling $T(\pmb{r})$ is local and can be expanded into Fourier components
\begin{equation}
\label{Eq_Tr}
T(\pmb{r}) = w_0 \sum\limits_{j=1}^3 e^{i \tilde{\pmb{g}}_j \cdot \pmb{r}} T_j,
\end{equation}
with $T_1 = \alpha \sigma_0 + \sigma_x$, $T_2 = \alpha \sigma_0 + \cos\phi \sigma_x + \sin\phi \sigma_y$ and $T_3 = \alpha \sigma_0 + \cos\phi \sigma_x + \sin\phi \sigma^*_y$. $\tilde{\pmb{g}}_j$ is related to three nearest-neighbour interlayer momentum shifts $\pmb{q}_j$ by $\tilde{\pmb{g}}_j = \pmb{q}_j + \pmb{K}_1 - \pmb{K}_2$.
The corrugation effects, effectively added to the BM model by setting $\alpha < 1$ \cite{TBGmodel_Koshino_2018}, separate the flat bands from remote bands. 
In the following calculations, we use same parameters as in Ref.~\cite{TBGmodel_Koshino_2018}:
\begin{equation}
\begin{split}
\hbar v_{\rm F}/a = 2135.4 {\rm meV}, w_0 = 79.7 {\rm meV}, \alpha = 0.82.
\end{split}
\end{equation}

\begin{figure*}[t]
\centering
\includegraphics[width=1.0\textwidth]{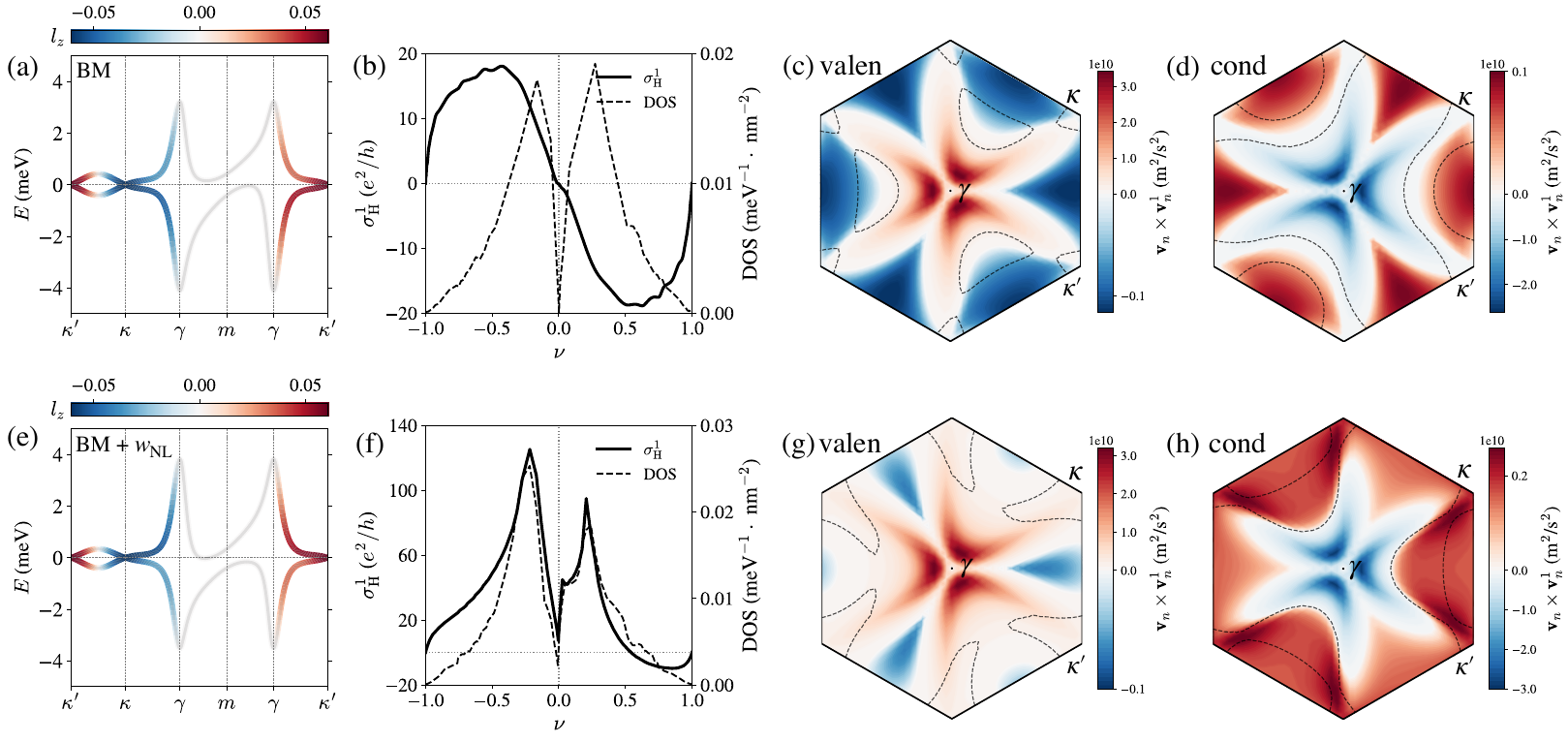}
  \vspace{-5pt}
\caption{\label{fig_BM} { 
From left to right: the MATBG ($1.05^\circ$) flat-band band structure, DOS and layer Hall conductivity $\sigma_{\rm H}^{1}$ as a function of one-flavor filling factor $\nu$, and $\pmb{v}_n \times \pmb{v}^{1}_n$ distribution in the first moir\'e Brillouin zone for valence and conduction bands.
(a-d) The BM model with local interlayer tunneling $T(\pmb{r})$.
(e-h) The generalized BM model with non-local interlayer tunneling strength $w_{\rm NL}=-10$. The band structures are colored by the layer polarization $l_z$. The black dashed lines in (c-d) and (g-h) mark the Fermi surface contours at DOS peaks in the valence and conduction flat bands: (c) at $\nu \approx -0.2$, (d) at $\nu \approx 0.3$, (g) at $\nu \approx -0.2$, (h) at $\nu \approx 0.2$.
  }}
\end{figure*}

\begin{figure*}[t]
\centering
\includegraphics[width=1.0\textwidth]{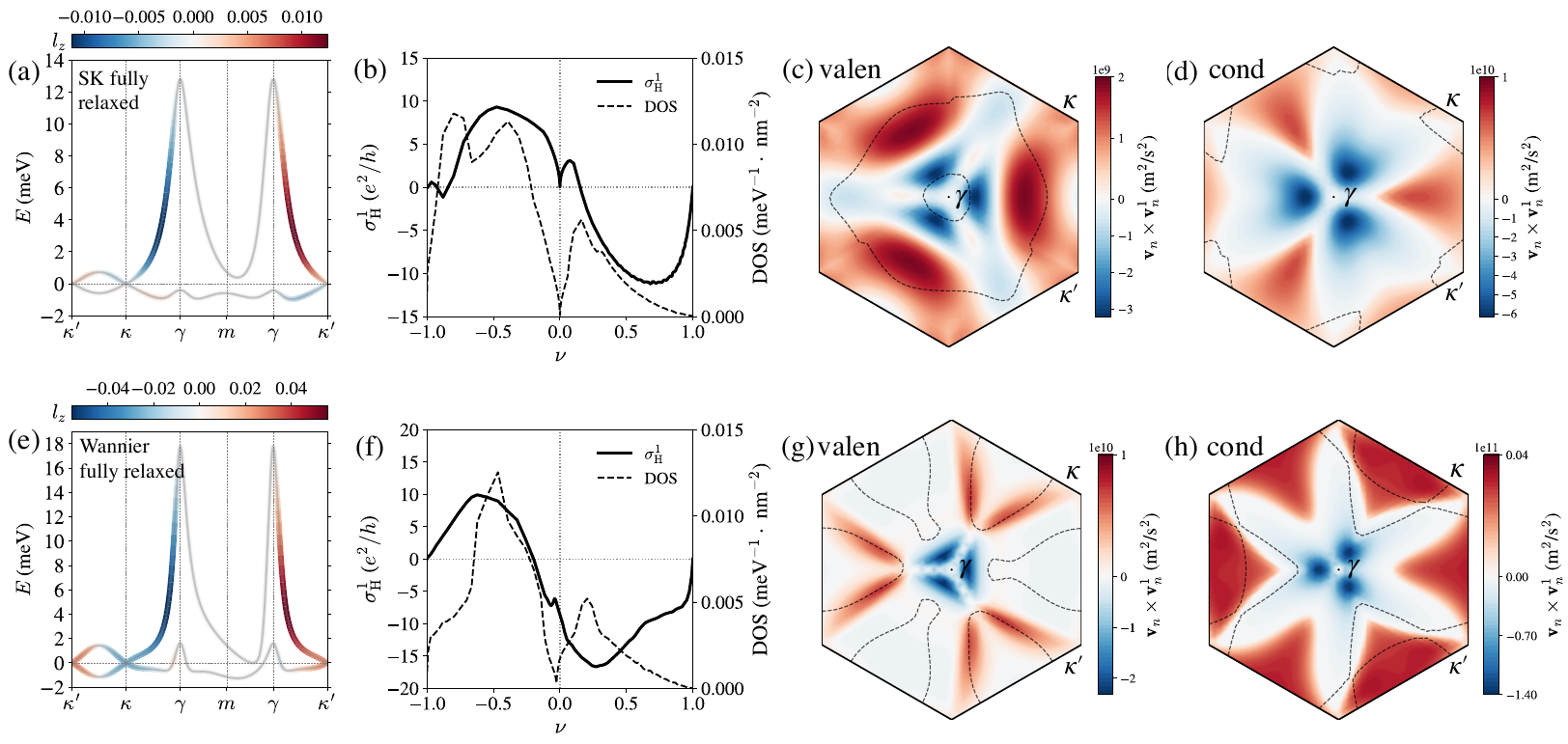}
  \vspace{-5pt}
\caption{\label{fig_relaxed} { 
Same figures as in Fig.~\ref{fig_BM}.
(a-d) The fully relaxed model using Slater-Koster tight-binding parameters.
(e-h) The fully relaxed model using Wannier tight-binding parameters.
The black dashed lines mark the Fermi surface contours near DOS peaks: (c) at $\nu \approx -0.45$, (d) at $\nu \approx 0.1$, (g) at $\nu \approx -0.5$, (h) at $\nu \approx 0.2$.
  }}
\end{figure*}

\subsection{The generalized BM model with non-local interlayer tunneling}
\label{sec_MATBGmodels_nonlocal}
The particle-hole symmetry breaking generally observed in experiments cannot be captured by the BM model described in the previous subsection. One way to incorporate
the particle-hole symmetry breaking is to add the non-local, i.e. the first-order gradient, interlayer tunneling term~\cite{Xie_WeakField_2021} to the BM model. Instead of the local interlayer tunneling Eq.~(\ref{Eq_Tr}), we use
\begin{equation}
T(\pmb{r}, \pmb{r}') = \frac{1}{A} \sum\limits_{\pmb{k}, \pmb{k}'} e^{i\pmb{k} \cdot \pmb{r}} e^{-i\pmb{k}' \cdot \pmb{r}'} T_{\pmb{k} \pmb{k}'},
\end{equation}
where
\begin{equation}
\label{Eq_nonlocalT}
\begin{split}
T_{\pmb{k} \pmb{k}'} 
&= \frac{1}{A_{\rm uc}} \sum\limits_{j=1}^3 \delta_{\pmb{k}-\pmb{k}', \tilde{\pmb{g}}_j} t(\pmb{k}+\pmb{G}_j) T_j \\
&= \sum\limits_{j=1}^3 \delta_{\pmb{k}-\pmb{k}', \tilde{\pmb{g}}_j} \big[ w_0 + \frac{w_{\rm NL}}{g_{\rm M}}(|\pmb{k}+\pmb{G}_j|-k_{\rm D}) \big] T_j.
\end{split}
\end{equation}
Only three nearest neighbour momentum transfers for the interlayer tunnelling are included.
The non-local interlayer tunneling strength is defined as
\begin{equation}
\begin{split}
w_{\rm NL} = \frac{g_{\rm M}}{A_{\rm uc}} \frac{dt}{dk} \Big|_{k=k_{\rm D}},
\end{split}
\end{equation}
where $g_{\rm M}$ is the length of moir\'e primitive reciprocal lattice vector, $A_{\rm uc}$ is the moir\'e unit cell area and
$dt/dk$ characterizes the rate at which the interlayer tunneling strength $t$ diminishes with momentum $k$ in the two-center approximation. $w_{\rm NL}$ is a tuning parameter in this model and in our calculations shown in the next section we choose it to be $w_{\rm NL}=-10$. The qualitative features of this generalized BM model that we will discuss later do not depend on 
the specific value of $w_{\rm NL}$.

\subsection{The comprehensive continuum model with relaxations}
\label{sec_MATBGmodels_relax}

References \cite{JKang_TBGmodel_2023, Vafek_TBGmodel_2023} constructed the effective continuum Hamiltonian of MATBG 
following a systematic derivation of the
real-space continuum model, which takes into account atomic relaxations. 
A thorough review can be found in Appendix~\ref{App:Model3}. Here, we outline its essential characteristics.
The Hamiltonian undergoes modifications through atomic relaxations in three ways:
(i) The in-plane relaxation is incorporated as a pseudo-gauge field, 
which couples different momentum states within the same layer; 
(ii) Both in-plane relaxation (which shrinks AA-stacking area) and out-of-plane relaxation (which increases the vertical atomic distance near AA-stacking) result in larger off-diagonal elements than the diagonal elements of the interlayer tunneling term. This correction is relevant for the band gap between the flat bands and remote bands. 
(iii) The formation of the relaxed domain wall enhances the scattering involving larger momentum states.

\section{Layer Hall conductivity using different model parameters}\label{sec_para}
This section investigates the layer Hall conductivity of MATBG ($\sim 1.05^\circ$) flat bands by referencing previously established formula.
We use different single-particle continuum models and parameters described in the earlier Sec.~\ref{sec_MATBGmodels} to draw comparisons.

We initiate our discussion by analyzing the layer Hall conductivity within the framework of the BM model in Sec.~\ref{sec_MATBGmodels_BM}
and the generalized BM model augmented by the inclusion of the non-local interlayer tunnelling term in Sec.~\ref{sec_MATBGmodels_nonlocal}.
In Fig.~\ref{fig_BM}, we show the flat band spectrum, density of states (DOS) and layer Hall conductivity $\sigma_{\rm H}^{1}$ of the BM model (Fig.~\ref{fig_BM}(a, b)) and of the generalized BM model with the non-local interlayer tunneling strength $w_{\rm NL}=-10$ (Fig.~\ref{fig_BM}(c, d)). The spectra are colored by the layer polarization $l_z$, which is the eigenvalue of the Pauli matrix $\hat{l}_z = {\rm diag}(1,-1)$ acting on the layer subspace. As expected, the states are strongly layer hybridized with $l_z\sim0$.
When comparing the band structures in Fig.~\ref{fig_BM}(a) and Fig.~\ref{fig_BM}(e), the non-local interlayer tunnelling has little effect on the spectrum, slightly uplifting the $\gamma$ point energies. 
In both cases the conductivity  shows most pronounced magnitude around regions with the largest DOS, which is expected as $\sigma_{\rm H}^{1}$ is a Fermi surface property.
Importantly, however, the non-local interlayer tunneling, in the form of Eq.(\ref{Eq_nonlocalT}), amplifies $\sigma_{\rm H}^{1}$ by an order of magnitude and flips the sign of $\sigma_{\rm H}^{1}$ when the system is doped with electrons, as shown in Fig.~\ref{fig_BM}(f) when contrasting with the results of the original BM model in Fig.~\ref{fig_BM}(b).
To better understand these differences in $\sigma_{\rm H}^{1}$ due to the non-local interlayer tunneling, we plot $k$-space distribution of $\pmb{v}_n \times \pmb{v}^{1}_n$ in Fig.~\ref{fig_BM}(c-d) for the original BM model with local $T(\pmb{r})$ and in Fig.~\ref{fig_BM}(g-h) for the generalized BM model with non-local $T(\pmb{r},\pmb{r}')$. 
Even though there is a resemblance in $\pmb{v}_n \times \pmb{v}^{1}_n$ distributions, the Fermi surface contours at DOS peaks differ in the two models, which are
marked in black dashed lines in these figures.
As $\sigma_{\rm H}^{1}$ is an integral of $\pmb{v}_n \times \pmb{v}^{1}_n$ over states near the Fermi surface, as shown in Eq.~(\ref{Eq_sigmaH_surface}), these distinct Fermi surface contours account for the amplified positive $\sigma_{\rm H}^{1}$ in the non-local tunneling model. In particular, $\sigma_{\rm H}^{1}$ aligns more closely with DOS in this model, as depicted in Fig.~\ref{fig_BM}(f).

In another comparison, we show same figures in Fig.~\ref{fig_relaxed} for the comprehensive continuum model that factors in relaxation in Sec.~\ref{sec_MATBGmodels_relax}, using the Slater-Koster tight-binding model \cite{Moon_SK_2012} parameters (Fig.~\ref{fig_relaxed}(a-d)) and the Wannier tight-binding model \cite{Fang_Wannier_2016} parameters (Fig.~\ref{fig_relaxed}(e-h)).
Compared to the BM models in Fig.~\ref{fig_BM}, the inclusion of full relaxations broadens the flat conduction band from 4 meV to 13-18 meV and narrows the flat valence band from 4 meV to 1 meV.
A direct consequence of the narrowed valence band is that, instead of a sharp peak in DOS in BM models, the DOS in Fig.~\ref{fig_relaxed}(b,f) is generally high for hole-doped fillings.
Despite relaxation effects significantly suppress the flat valence band bandwidth, the shapes and values of $\sigma_{\rm H}^{1}$ in relation to $\nu$ (Fig.\ref{fig_relaxed}(b)) remain consistent with the BM model in Fig.\ref{fig_BM}(b).
One notable electronic feature due to relaxation is a bump near the valence band $\gamma$ point, which reverses the sign of velocity, as shown in Fig.~\ref{fig_relaxed}(c,g). 
Furthermore, with Wannier tight-binding model parameters in Fig.~\ref{fig_relaxed}(e), the bump at $\gamma$ point becomes the global maximum of the valence band, resulting in hole-like states at charge neutrality. 
This characteristic trait in the band structure contributes directly to the non-zero $\sigma_{\rm H}^{1}$ at charge neutrality, as shown in Fig.~\ref{fig_relaxed}(f), once the Fermi level aligns with the valence band $\gamma$-point energy. This is in contrast to the feature in Fig.~\ref{fig_BM}(b,f) and Fig.~\ref{fig_relaxed}(b) that $\sigma_{\rm H}^{1}=0$ at $\nu=0$, a result of Dirac-point physics.
For the conduction band, the $\pmb{v}_n \times \pmb{v}_n^{1}$
distribution bears a resemblance to the BM model excecpt that the distribution patterns are rotated by $180^\circ$, as seen in Fig.~\ref{fig_relaxed}(d,h) and Fig.~\ref{fig_BM}(d,h).
Generally, the influences observed in the Slater-Koster and Wannier models are qualitatively alike. 

In addition to the Dirac term $h_{\rm D}$ and the local interlayer tunneling $T$ present in the BM model in Eq.~(\ref{Eq_H_BM}), the fully relaxed continuum model introduces several other terms. We discuss the effects of each individual term in the Appendix~\ref{App:Model3Individual}.

\begin{figure*}[t]
\centering
\includegraphics[width=1.0\textwidth]{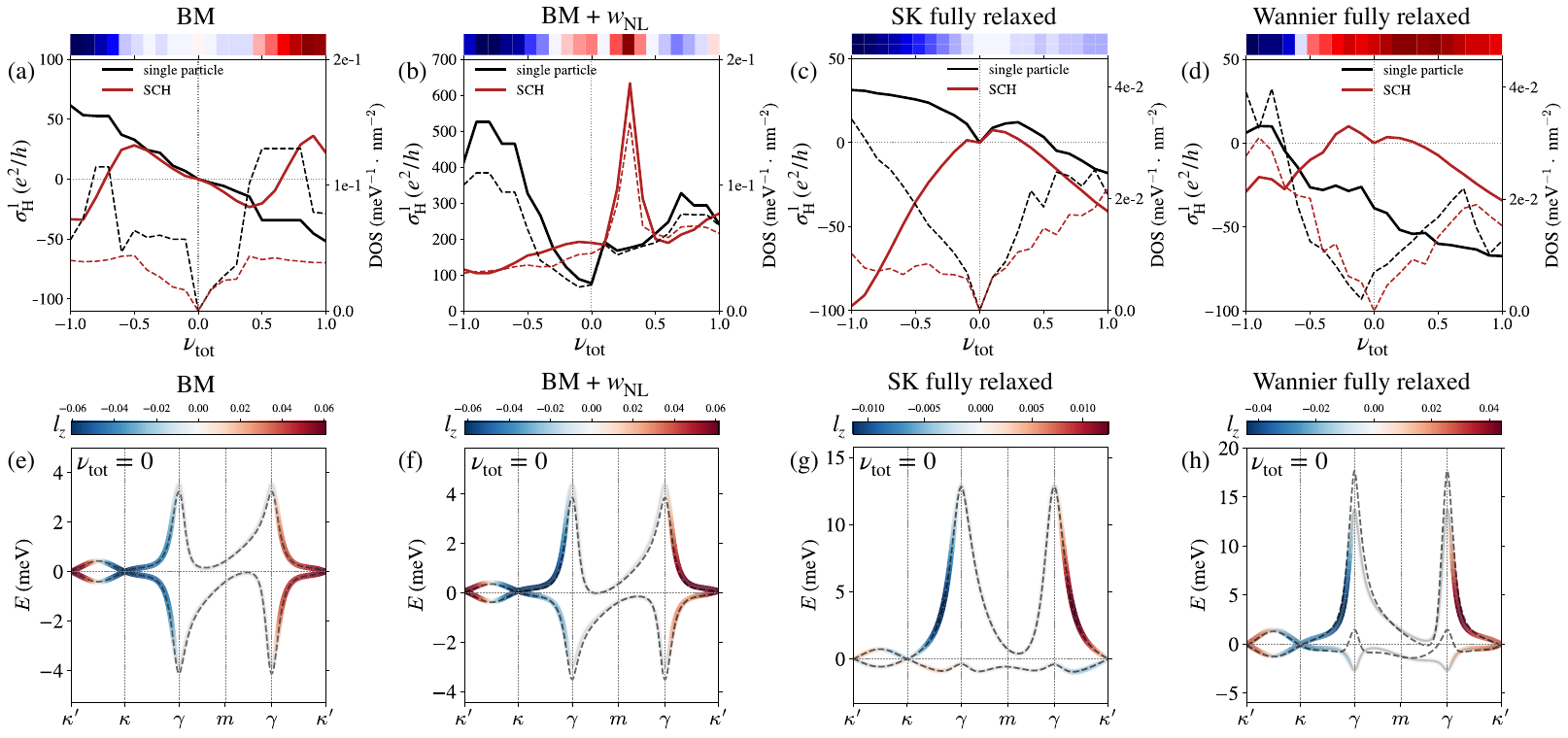}
  \vspace{-5pt}
\caption{\label{fig_SCH_nu0} { 
(a-d)
$\sigma_{\rm H}^{1}$ (solid lines) and DOS (dashed lines) versus the total filling factor $\nu_{\rm tot}$ including four flavors of the SCH quasiparticle bands calculated based on (a) the BM model, (b) the genralized BM model with non-local interlayer tunneling $w_{\rm NL}=-10$, (c) the fully relaxed model with SK model parameters and (d) the fully relaxed model with Wannier model parameters. Note that we only show SCH calculations for $\nu_{\rm tot} \in [-1, 1]$, within which exchange interactions are canceled by correlations such that the TR symmetry is not spontaneously broken. The colorbar on top of each figure illustrates the layer Hall conductivity difference between the SCH and single-particle calculations (red/blue: increase/decrease). Four single-particle continuum models have very different SCH responses.
(e-h) SCH (colored spectra) and single-particle (dashed curves) band structures at $\nu_{\rm tot}=0$.
}}
\end{figure*}

\begin{figure*}[t]
\centering
\includegraphics[width=1.0\textwidth]{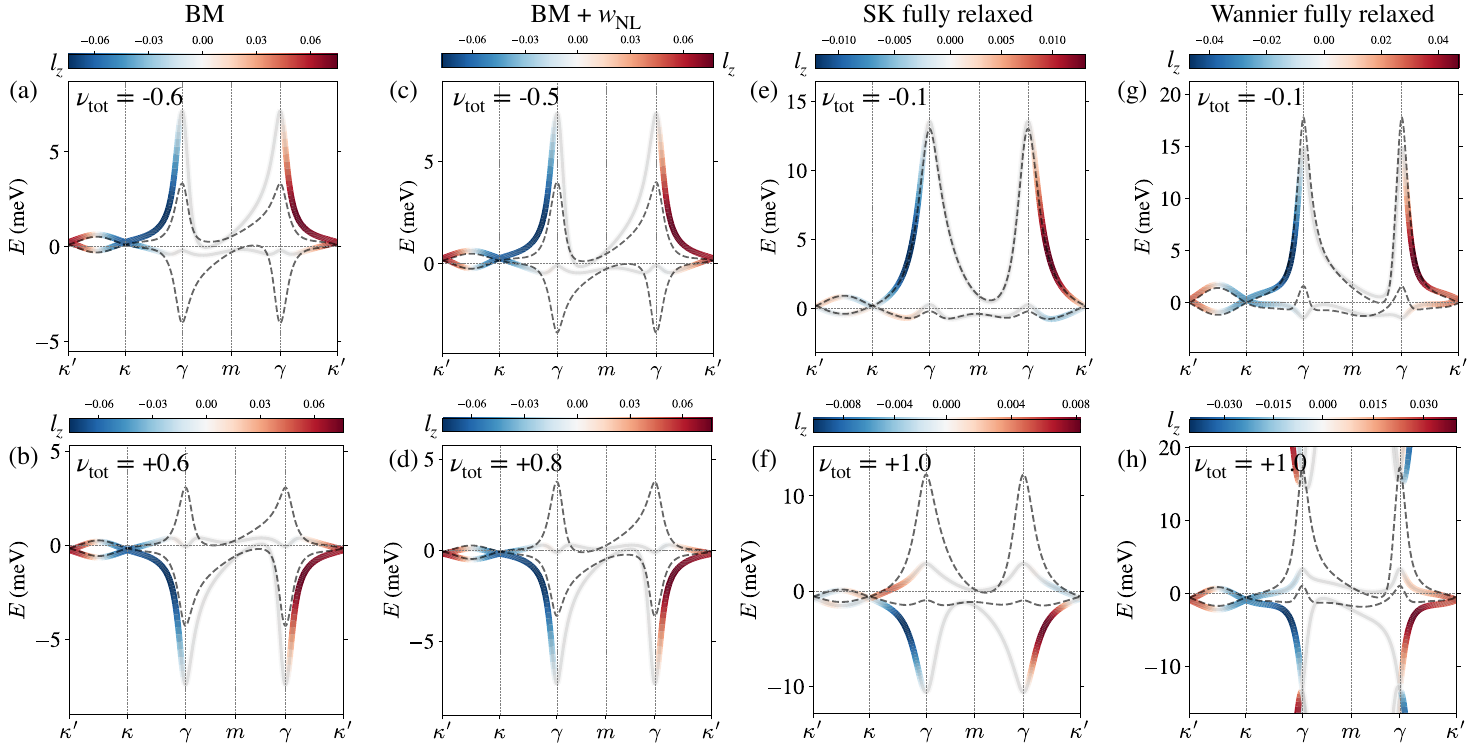}
  \vspace{-5pt}
\caption{\label{fig_SCH_nu} { 
(e-h) SCH (colored spectra) and single-particle (dashed curves) band structures at several chosen $\nu_{\rm tot}$.
}}
\end{figure*}

\section{Layer Hall conductivity of SCH quasiparticle bands}\label{sec_sch}
From the findings in the preceding section, the relaxation effects at the single-particle level, profoundly influence MATBG flat band structures and the associated transport characteristics reflected by the layer Hall conductivity. 
The extent of this influence is highly sensitive to the specific MATBG single-particle models and their parameters. 
This sensitivity complicates the task of accurately modeling MATBG and raises the question of which single-particle model should be chosen as the starting point.
Moreover, it challenges our ability to provide explanations or predictions corresponding to related experiments.

Experimental observations, especially in systems with flat bands, encompass all many-body effects which must be thoroughly considered.
In MATBG, the exchange interactions has emerged as crucial, leading to spontaneous TR-symmetry breaking, reminiscent of the Stoner model. 
Spin-valley flavor polarized insulating and metallic states have been identified near $\nu_{\rm tot} = 3$ (where $\nu_{\rm tot}$ represents the total filling factor, accounting for four spin-valley flavors) in multiple MATBG samples \cite{MagnetizationSwitching1, MagnetizationSwitching2}.
This reveals the symmetry breaking brought about by exchange interactions. 
For other samples, however, the spin-valley flavor symmetry remains intact within the filling range $\nu_{\rm tot} \in (-2,1)$ \cite{zondiner2020cascade, wong2020cascade, Xie_WeakField_2021}. 
At these small fillings approaching the charge neutrality, the influence of correlation effects intensifies, against the tendency of symmetry breaking by exchange interactions.
In MATBG, correlations could potentially and almost completely negate the impact of exchange effects for $\nu_{\rm tot} \in (-2,1)$ \cite{JZhu_GW_2023}. As a result, the electrostatic Hartree potential ends to be the predominant factor driving band renormalizations.
In general, the repulsive Hartree interaction in MATBG consistently elevates the energies of $\kappa$ and $\kappa'$ when electrons are added to the moiré unit cell one by one, keeping the energies at $\gamma$ relatively constant.

Accepting the perspective mentioned above, we posit that the MATBG ground state exhibits flavor paramagnetism at fillings close to the charge neutrality. This implies that all four spin-valley flavors are populated equally.
With this premise, we investigate the layer Hall conductivity of the SCH quasiparticle bands for $\nu_{\rm tot} \in [-1, 1]$ in this section.
The SCH calculations are conducted based on the continuum models previously outlined in Sec.~\ref{sec_MATBGmodels}.

We present our main results in Fig.~\ref{fig_SCH_nu0}(a-d), which shows the DOS (dashed lines) and $\sigma_{\rm H}^{1}$ (solid lines) as a function of $\nu_{\rm tot} \in [-1,1]$.
The red (black) lines represent calculations using the SCH (single-particle) bands.
The changes in $\sigma_{\rm H}^{1}$ due to the SCH potential are represented by the colorbar on top of each figure.
Despite shared features, the layer Hall responses of the SCH renormalized bands in these four models vary significantly, which will be further elaborated below.

First we discuss the results around the charge neutrality.
In Fig.~\ref{fig_SCH_nu0}(e-h) we show SCH-renormalized bands at $\nu_{\rm tot}=0$, where the electrostatic Hartree energy is often considered negligible due to the uniform density. 
As illustrated in Fig.~\ref{fig_SCH_nu0}(e,g), the SCH-renormalized bands (depicted by the colored spectrum) closely align with the single-particle band structures (dashed curves) in the BM model and the fully relaxed SK model. Accordingly, both $\sigma_{\rm H}^{1}$ and DOS maintain behavior analogous to the single-particle model near $\nu_{\rm tot}=0$, as depicted in Fig.~\ref{fig_SCH_nu0}(a,c). However, in the generalized BM model with the non-local interlayer tunneling and the fully relaxed Wannier model, band structures are renormalized—slightly in the former (Fig.~\ref{fig_SCH_nu0}(f)) and observably in the latter (Fig.~\ref{fig_SCH_nu0}(h))—even at charge neutrality. 
This band renormalization causes $\sigma_{\rm H}^{1}$ and DOS to deviate from the single-particle model, evident in Fig.~\ref{fig_SCH_nu0}(b,d). Particularly, in the fully relaxed Wannier model (Fig.~\ref{fig_SCH_nu0}(d)), the SCH potential lowers the energies at $\gamma$, eliminating the bump at $\gamma$ and yielding a zero value for both $\sigma_{\rm H}^{1}$ and DOS at charge neutrality.

At finite dopings, we emphasize the distinctions between different models.
For the BM model in Fig.~\ref{fig_SCH_nu0}(a), both $\sigma_{\rm H}^{1}$ and DOS of SCH bands follow the trends of the single-particle model for $\nu_{\rm tot} \in (-0.5,0.5)$. However, SCH effects suppress the DOS peaks and tend to reverse the sign of $\sigma_{\rm H}^{1}$ for $|\nu_{\rm tot}| \gtrsim 0.5$.
This can be explained by observing in Fig.~\ref{fig_SCH_nu}(a,b) that, SCH bands maintain the Dirac cone characteristics at small $|\nu_{\rm tot}|$, while velocities near $\gamma$ starts to reverse its sign for $|\nu_{\rm tot}| > 0.5$.
Unique to the BM model, $\sigma_{\rm H}^{1}$ behaves as an odd function of $\nu_{\rm tot}$ in Fig.~\ref{fig_SCH_nu0}(a).

In the generalized BM model with the non-local interlayer tunneling (Fig.~\ref{fig_SCH_nu0}(b)), 
the band renormalizations by SCH potential as a function of filling
exhibit similarities to the BM model, as shown in Fig.~\ref{fig_SCH_nu}(c,d). 
Notably, $\sigma_{\rm H}^{1}$ is positive across the filling range with much larger magnitude, which is a unique characteristic of the generalized BM model. Furthermore, it closely aligns with DOS, emphasizing DOS's pivotal role in determining the layer Hall conductivity.

In the two fully relaxed models, as shown in Fig.\ref{fig_SCH_nu0}(c,d), the conductivity remains overall negative within the considered doping range, which distinguishes them from the two BM models. 
Other subtle differences can also be identified \footnote{Another feature observed in the relaxed SK model is that the SCH potential with electron doping (Fig.~\ref{fig_SCH_nu}(f)) switches the layer polarization near $\kappa$ and $\kappa'$.}. 
For example, in contrast to the BM models (e.g., Fig.~\ref{fig_SCH_nu}(b, d)), the velocities near conduction band $\gamma$-point do not change sign with electron doping in the fully relaxed models (e.g., Fig.~\ref{fig_SCH_nu}(f,h)). This results in the increase of the layer Hall conductivity when $\nu_{\rm tot} \gtrsim 0.5$ in Fig.~\ref{fig_SCH_nu0}(a,b), while $\sigma_{\rm H}^{1}$ depicted in Fig.~\ref{fig_SCH_nu0}(c, d) continues to decline as electrons are added.

The two relaxed models can be distinguished by the distinct behaviors of the layer Hall conductivity when holes are introduced.
In the relaxed SK model, the valence band energy at $\gamma$ crosses the Fermi level immediately with hole doping, as seen in Fig.~\ref{fig_SCH_nu}(e). This results in a velocity in the opposite direction, causing a consistent decrease in $\sigma_{\rm H}^{1}$ for $\nu_{\rm tot} < 0$, which can be observed in Fig.~\ref{fig_SCH_nu0}(c).
This behavior, wherein $\sigma_{\rm H}^{1}$ drops on both the electron and hole doping sides, accounts for the overall negative $\sigma_{\rm H}^{1}$ seen in Fig.~\ref{fig_SCH_nu0}(c). 
Conversely, in the fully relaxed Wannier model, a slight hole doping reduces the $\gamma$-point energies, as illustrated in Fig.~\ref{fig_SCH_nu}(g). This contrasts with the other three models and explains why the layer Hall conductivity for the relaxed Wannier model is negative but less pronounced on the hole-doping side than in the SK model, as displayed in Fig.~\ref{fig_SCH_nu0}(d).

\section{Discussions}\label{sec_discussion}
It is well-recognized that MATBG properties, including phenomena like superconductivity and magnetism, exhibit significant sample variability, making it challenging to accurately model each MATBG sample that aligns with or anticipates experimental results.
In this paper, we provide a comprehensive comparison of the electronic band structure and layer Hall conductivity--which is influenced by factors like DOS and layer hybridization--across three different MATBG continuum models.
Even though commonly used as starting frameworks for many-body interacting models, these continuum models already exhibit distinct electronic and transport properties even at the single-particle level.

In MATBG, electrostatic Hartree potentials are important at small filling factors near the charge neutrality, where exchange interactions are largely offset by correlations. 
We, therefore, further explore the impact of SCH potentials on flat band dispersion and layer Hall conductivity.
Our calculations show that SCH potentials substantially renormalize the flat band dispersion and hence the layer Hall conductivity--the renormalization details depend on the specific continuum model used. 
The SCH-renormalized layer Hall conductivity, calculated based on the three continuum models, show different characteristics which can distinguish between these models. 
Given these findings, we propose that the layer Hall conductivity can serve as a viable experimental observable for distinguishing among these MATBG continuum models, even when factoring in SCH band renormalizations.

Another significant discovery from our study is that the layer Hall conductivity is particularly large in all the models, reaching $O(100)e^2/h$ with a moderate relaxation time of $\tau=1$ ps. Especially, in the generalized BM model with non-local interlayer tunneling, we find $\sigma_{\rm H}^{1} \sim 700 e^2/h$ as illustrated in Fig.~\ref{fig_SCH_nu0}(b), which is nearly a ten-fold increase relative to other MATBG continuum models that we consider in this paper. 
Large layer Hall conductivity implies a significant in-plane magnetic moment density induced by a DC electric field $\bs{m}_{\parallel}=-d_0\sigma_{\rm H}^{1}\bs{E}$ (Fig.~\ref{fig_m}).
For MATBG, we estimate $m_{\parallel}\approx -0.1 \mu_B/\rm nm^2$ ($2 \mu_B$ per moir\'e unit cell) with $d_0=0.335$ nm, $E=10^{4}$ V/m, $\tau=10$ ps \cite{He2020giant}, and $\sigma_{\rm H}^{1}=7000e^2/h$, where $\mu_B=e\hbar/2m$ is the Bohr magneton and $m$ is the free electron mass. Such a large in-plane magnetoelectric response is even greater than the “giant” out-of-plane one in strained TBG \cite{He2020giant}, and is comparable to the equilibrium magnetization in MATBG. Previous experiments have shown that a current can switch the direction of magnetization in MATBG~\cite{MagnetizationSwitching1,MagnetizationSwitching2}. The induced sizable $\bs{m}_{\parallel}$ discussed in our work can in principle provide the in-plane component that is necessary during the magnetic switching process.
Around the magic angle, one might expect that the CD could also be observable even though the structural chirality is weak due to a small twist angle.

In this work, we have specifically concentrated on the TR-even charge Hall conductivity contributed equally by the two valleys.
In SCH calculations, our focus has been on the total filling factor range $\nu_{\rm tot} \in [-1, 1]$, where the TR symmetry is often approximately preserved in MATBG. 
At higher electron or hole dopings, TR symmetry broken states can spontaneously emerge as a result of exchange interactions, rendering an intrinsic charge Hall current contributed by the anomalous velocity.
Note that such intrinsic charge Hall currents are distinct in the two layers of a chiral structure, leading to an in-plane magnetic moment $\bs{m}_{\parallel}$. It would be interesting to study such intrinsic $\bs{m}_{\parallel}$ in future studies. The CD of MATBG in the presence of interactions is another interesting aspect to explore, which could potentially be measured with terahertz techniques.

\begin{acknowledgments}
We thank Allan H. MacDonald for insightful discussions and Chengxin Xiao for the assistance of preparing Fig.~\ref{fig_m}. DZ and WY were supported by National Key R\&D Program of China (2020YFA0309600), and Research Grant Council of HKSAR (AoE/P-701/20, HKU SRFS2122-7S05). WY also acknowledges support by New Cornerstone Science Foundation. 
\end{acknowledgments}


%

\vspace{0em} 
\appendix
\section{A review of the comprehensive continuum model with relaxations}\label{App:Model3}

At a specific valley $K$, the relaxed Hamiltonian consists of intralayer and interlayer terms,
$\mathcal{H}_{K} = \mathcal{H}_{\rm intra} + \mathcal{H}_{\rm inter}$. Up to $\sim 1$ meV accuracy in energy compared with microscopic tight-binding models, the intralayer (interlayer) Hamiltonian is expanded up to the second (first) order in gradients. From Ref.~\cite{JKang_TBGmodel_2023},
the intralayer Hamiltonian is
\begin{widetext}
\begin{equation}
\label{Eq_Hintra}
\begin{split}
\mathcal{H}_{\rm intra} = \int d^2 \pmb{r} \sum\limits_{j=t,b} \sum\limits_{SS'} \Psi^\dagger_{j,S}(\pmb{r}) \Bigg\{ \Big[ \mu + \beta_0 p^2 + \frac{C_0}{2}  \big( \pmb{p} \cdot \pmb{A}(\pmb{r}) + \pmb{A}(\pmb{r}) \cdot \pmb{p} \big) \Big] \delta_{SS'} + v_F \bs{\sigma}_{SS'} \cdot \big[ \pmb{p}^{(j)} + \gamma \pmb{A}^{(j)}(\pmb{r}) \big] \\
+ \beta_1 \big[ (p_x^2 - p_y^2) \sigma_x - 2p_x p_y \sigma_y \big]_{SS'} + \frac{1}{2} \sum\limits_\mu \big[ p_\mu \xi_{\mu,SS'}(\pmb{r}) + \xi_{\mu,SS'}(\pmb{r}) p_\mu \big] \Bigg\} \Psi_{j,S'}(\pmb{r}),
\end{split}
\end{equation}
\end{widetext}
where $j$ is summed over top (t) and bottom (b) layers, $S,S'$ are summed over A and B sublattices, and $\mu$ over $x$ and $y$ components. 
Because of the rotated Dirac points, $\pmb{p}^{(j)}$ is the operator $\pmb{p} = -i\hbar \nabla = -i\hbar \partial/\partial \pmb{r}$ rotated by $\mp \theta/2$.
For $|\theta| \ll 1$,
\begin{equation}
\begin{split}
\pmb{p}^{(j)} \approx (p_x + \theta_j p_y, p_y - \theta_j p_x),
\end{split}
\end{equation}
where top (bottom) layer is rotated anticlockwise (clockwise), $\theta_t = -\theta_b= \theta/2$.

The lattice-distortion induced pseudovector fields $\pmb{A}(\pmb{r})$ and $\pmb{\xi}(\pmb{r})$ are periodic with moiré periodicity,
\begin{equation}
\begin{split}
\pmb{A}(\pmb{r}) &= \sum\limits_{\pmb{g}} \pmb{A}_{\pmb{g}} e^{i \pmb{g} \cdot \pmb{r}}, \quad
\pmb{\xi}(\pmb{r}) = \sum\limits_{\pmb{g}} \pmb{\xi}_{\pmb{g}} e^{i \pmb{g} \cdot \pmb{r}} 
\end{split}
\end{equation}
with Fourier components 
\begin{equation}
\begin{split}
\pmb{A}_{\pmb{g}} &= \big(-2 g_x g_y , \ -(g_x^2-g_y^2) \big) \tilde{\varepsilon}^U_{\pmb{g}}
\end{split}
\end{equation}
and
\begin{widetext}
\begin{equation}
\begin{split}
\xi_{\pmb{g},x} &= \Big\{ -\Big(\frac{v_F}{2}+2 D_0 \Big) g_x g_y \sigma_x - \Big[ \Big( \frac{v_F}{2}+D_0 \Big) g_y^2 - D_0 g_x^2 \Big] \sigma_y \Big\} \tilde{\varepsilon}^U_{\pmb{g}},
 \\
\xi_{\pmb{g},y} &= \Big\{ \Big[ \Big( \frac{v_F}{2}+D_0 \Big) g_x^2 - D_0 g_y^2 \Big] \sigma_x + \Big(\frac{v_F}{2}+2 D_0 \Big) g_x g_y \sigma_y  \Big\} \tilde{\varepsilon}^U_{\pmb{g}}.
\end{split}
\end{equation}
\end{widetext}
$\tilde{\varepsilon}^U_{\pmb{g}}$ are Fourier components of the scalar field $\varepsilon^U(\pmb{r})$, which is the solenoidal part of the lattice relaxation $\delta U(\pmb{r}) \approx \pmb{\nabla} \times (\hat{\pmb{z}} \varepsilon^U(\pmb{r}))$. 
$\pmb{A}^{(j)}(\pmb{r})$ 
are rotated pseudovector fields defined as
\begin{equation}
\begin{split}
\pmb{A}^{(t)}(\pmb{r})
&= \mathcal{R}_{\theta_{t}} \pmb{A}
\approx (A_x - \theta_{t} A_y, A_y + \theta_t A_x ), \\
\pmb{A}^{(b)}(\pmb{r})
&= -\mathcal{R}_{\theta_{b}} \pmb{A}
\approx -(A_x - \theta_{b} A_y, A_y + \theta_b A_x ).
\end{split}
\end{equation}
The pseudovector fields for top and bottom layers are opposite in sign.

The interlayer part is written as
\begin{widetext}
\begin{equation}
\label{Eq_Hinter}
\begin{split}
\mathcal{H}_{\rm inter} = \sum\limits_{SS'} \int d^2 \pmb{r} \Psi^\dagger_{t,S}(\pmb{r}) \Big[ T_{SS'}(\pmb{r}) + \frac{1}{2} \{ \pmb{p}, \bs{\Lambda}_{SS'}(\pmb{r}) \} \Big] \Psi_{b,S'}(\pmb{r}) + h.c.
\end{split}
\end{equation}
\end{widetext}
$T(\pmb{r})$ and $\Lambda(\pmb{r})$ are local and non-local terms respectively.

In momentum space,
\begin{widetext}
\begin{equation}
\begin{split}
\mathcal{H}_{\rm intra} = \sum\limits_{\pmb{k},\pmb{k}',l} & c^\dagger_{\pmb{k},l} c_{\pmb{k}', l} \Big[ h_D^{\theta_l}(\pmb{k}-\pmb{K}_l) + \beta_0 |\pmb{k}-\pmb{K}_l|^2 \sigma_0 + \bar{\beta}_1(\pmb{k}-\pmb{K}_l) \Big] \delta_{\pmb{k} \pmb{k}'} \\
+ & c^\dagger_{\pmb{k},l} c_{\pmb{k}', l} \Big[ v_F \gamma \pmb{\sigma} \cdot \pmb{A}^{(l)}_{\pmb{g}} + \big( (\pmb{k}+\pmb{k}')/2 - \pmb{K}_l \big) \cdot \big(C_0 \pmb{A}_{\pmb{g}} \sigma_0 + \eta_{1,\pmb{g}} \sigma_x + \eta_{2,\pmb{g}} \sigma_y \big) \Big] \delta_{\pmb{k}-\pmb{k}', \pmb{g}}, \\
\end{split}
\end{equation}
\end{widetext}
and
\begin{widetext}
\begin{equation}
\begin{split}
\mathcal{H}_{\rm inter} = \sum\limits_{\pmb{k},\pmb{k}'} & c^\dagger_{\pmb{k},t} c_{\pmb{k}', b} \Big[ 
T_{\pmb{g}} + \frac{1}{2}(\pmb{k}+\pmb{k}'-\pmb{K}_t - \pmb{K}_b) \cdot \Lambda_{\pmb{g}}
 \Big] \delta_{\pmb{k}-\pmb{k}',\pmb{g}} + h.c.
\end{split}
\end{equation}
\end{widetext}
where
\begin{gather}
\begin{split}
&\bar{\beta}_1(\pmb{k}) = \beta_1\big[ (k_x^2-k_y^2) \sigma_x - 2k_x k_y \sigma_y \big], \\
&\pmb{\eta}_{1,\pmb{g}} = \Big( -\big(\frac{v_F}{2}+2 D_0 \big) g_x g_y, \big( \frac{v_F}{2}+D_0 \big) g_x^2 - D_0 g_y^2 \Big) \tilde{\varepsilon}^U_{\pmb{g}}, \\
&\pmb{\eta}_{2,\pmb{g}} = \Big( -\big(\frac{v_F}{2}+ D_0 \big) g_y^2+ D_0 g_x^2, \big( \frac{v_F}{2}+2D_0 \big) g_x g_y \Big) \tilde{\varepsilon}^U_{\pmb{g}}. 
\end{split}
\end{gather}
All model parameters in Eq.~(\ref{Eq_Hintra}) and Eq.~(\ref{Eq_Hinter}) are listed in Table.III and Table.V of Ref.~\cite{JKang_TBGmodel_2023}.


Hamiltonian $\mathcal{H}_K$ preserves the $\mathcal{C}_{3z}$ rotation, $\mathcal{C}_{2z}\mathcal{T}$, $\mathcal{C}_{2x}$ and $\mathcal{M}_z$ symmetries. $\mathcal{C}_{2x}$ is a $180^\circ$ rotation with respect to $x$-axis and swaps both layer and sublattice. $\mathcal{M}_z$ is the $z \rightarrow -z$ mirror which swaps layers.
Symmetry requirements on $A^{(j)}(\pmb{r})$ and $T(\pmb{r})$ are
\begin{equation}
\begin{split}
\mathcal{C}_{3z}:
& e^{i \sigma_z 2\pi/3} A^{(j)}(\mathcal{R}_3^{-1} \pmb{r}) e^{-i \sigma_z 2\pi/3} = A^{(j)}(\pmb{r}), \\
& e^{i \sigma_z 2\pi/3} T(\mathcal{R}_3^{-1} \pmb{r}) e^{-i \sigma_z 2\pi/3} = T(\pmb{r}),\\
\mathcal{C}_{2z}\mathcal{T}:
& \sigma_x A^{(j)*}(-\pmb{r}) \sigma_x = A^{(j)}(\pmb{r}),\\
& \sigma_x T^*(-\pmb{r}) \sigma_x = T(\pmb{r}), \\
\mathcal{C}_{2x}:
& \sigma_x A^{(1)}(\mathcal{M}_y \pmb{r}) \sigma_x  = A^{(2)}(\pmb{r}),\\
& \sigma_x T(\mathcal{M}_y \pmb{r}) \sigma_x = T^\dagger(\pmb{r}),\\  
\mathcal{M}_z:
& A^{(1)}(-\pmb{r}) = A^{(2)}(\pmb{r}), \\
& T(-\pmb{r}) = T^\dagger(\pmb{r}),\\
\end{split}
\end{equation}
and on their Fourier components
\begin{equation}
\begin{split}
\mathcal{C}_{3z}:
& e^{i \sigma_z 2\pi/3} A^{(j)}_{\mathcal{R}_3^{-1} \pmb{g}} e^{-i \sigma_z 2\pi/3} = A^{(j)}_{\pmb{g}}, \\
& e^{i \sigma_z 2\pi/3} T_{\mathcal{R}_3^{-1} \pmb{g}} e^{-i \sigma_z 2\pi/3} = T_{\pmb{g}}, \\
\mathcal{C}_{2z}\mathcal{T}:
& \sigma_x A_{\pmb{g}}^{(j)*} \sigma_x = A^{(j)}_{\pmb{g}}, \\
& \sigma_x T^*_{\pmb{g}} \sigma_x = T_{\pmb{g}}, \\
\mathcal{C}_{2x}:
& \sigma_x A^{(1)}_{\mathcal{M}_y \pmb{g}} \sigma_x  = A^{(2)}_{\pmb{g}}, \\
& \sigma_x T_{-\mathcal{M}_y \pmb{g}} \sigma_x = T^\dagger_{\pmb{g}}, \\
\mathcal{M}_z:
& A^{(1)}_{-\pmb{g}} = A^{(2)}_{\pmb{g}},\\
& T_{-\pmb{g}} = T^\dagger_{\pmb{g}}.
\end{split}
\end{equation}

In summary, atomic relaxations modify $\mathcal{H}_K$ in terms of three aspects: 
i) The in-plane relaxation is incorporated as the pseudo-gauge field $\pmb{A}^{(j)}(\pmb{r})$, which couples different momentum states within the same layer; 
ii) Both in-plane relaxation (which shrinks AA-stacking area) and out-of-plane relaxation (which increases the vertical atomic distance near AA-stacking) result in larger off-diagonal elements compared to the diagonal elements of $T(\pmb{r})$. This correction is relevant for the band gap between the flat bands and remote bands. 
iii) The formation of the relaxed domain wall enhances the scattering involving larger momentum states.

\section{Effects of the individual terms of the third model}\label{App:Model3Individual}

\begin{figure*}[!h]
\centering
\includegraphics[width=1\textwidth]{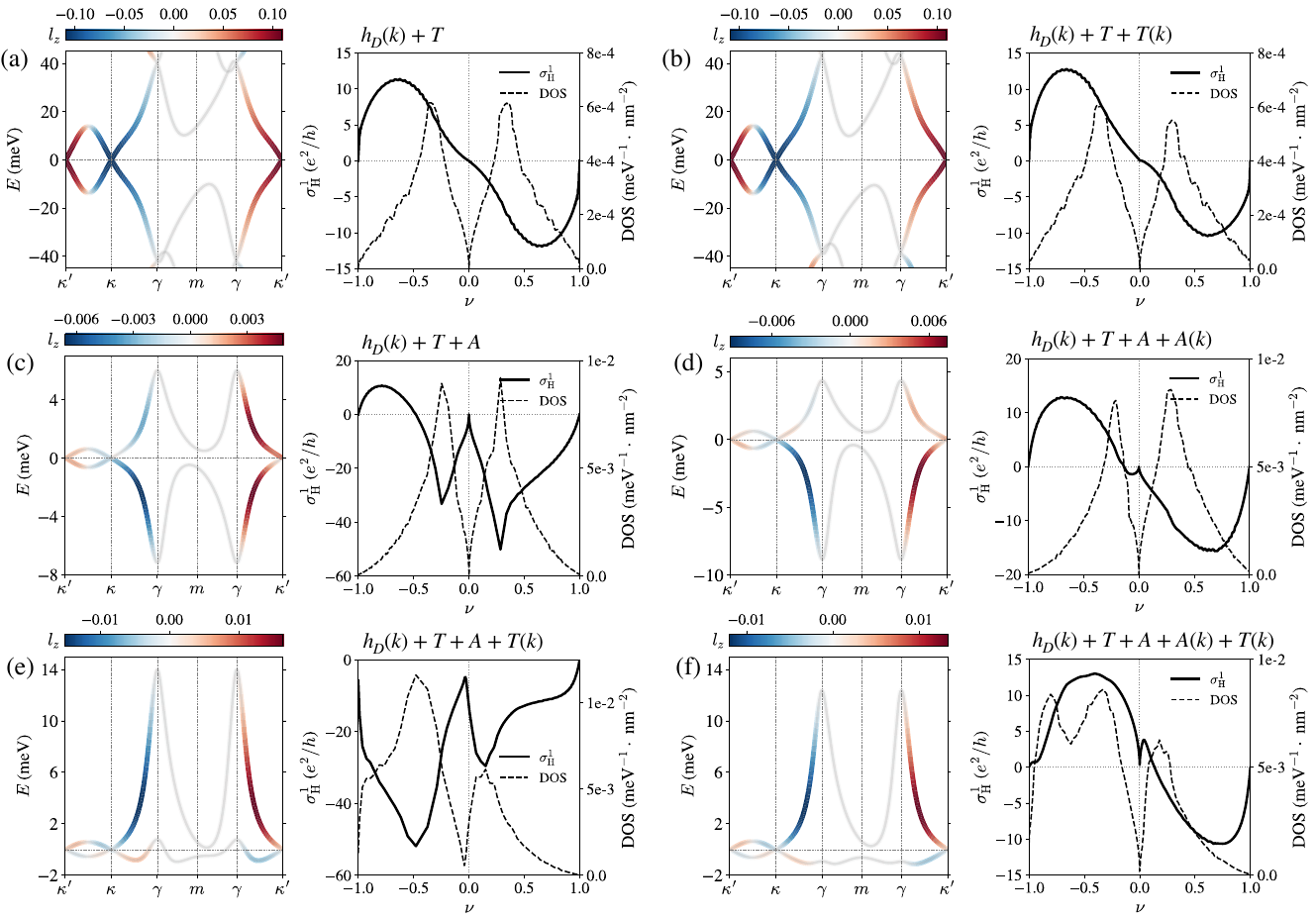}
  \vspace{-5pt}
\caption{\label{fig_SK} { 
Detailed comparison of the fully relaxed model using Slater-Koster tight-binding model parameters by individually deactivating $A$, $A(k)$, $T(k)$ and $\beta(k^2)$ terms in Eq.~(\ref{Eq_H_Vafek}). The effects of different terms are summarized in Tab.\ref{Tab_SK}.
}}
\end{figure*}

\begin{figure*}[!h]
\centering
\includegraphics[width=1\textwidth]{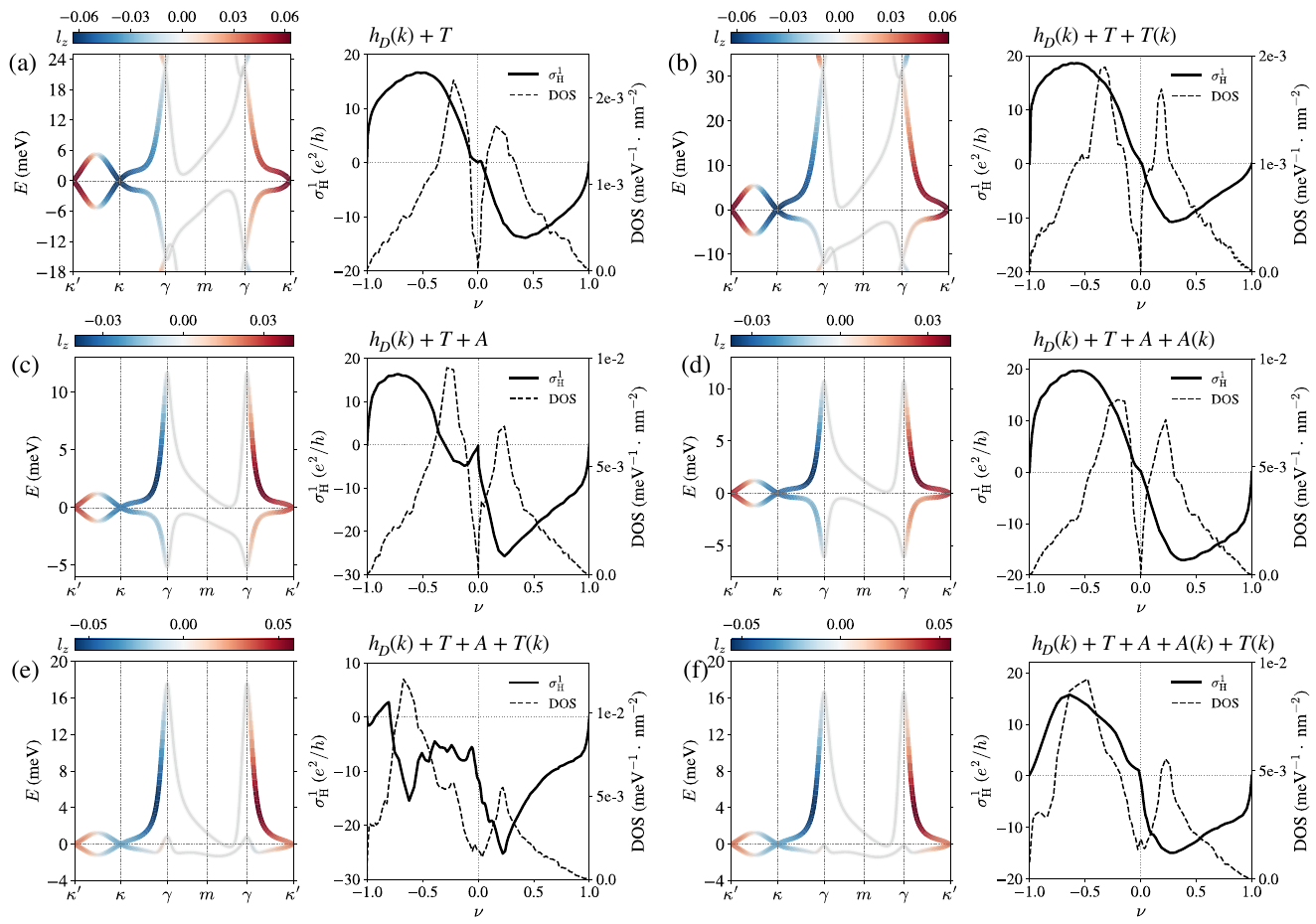}
  \vspace{-5pt}
\caption{\label{fig_Wan} { 
Detailed comparison of the fully relaxed model using Wannier tight-binding model parameters by individually deactivating $A$, $A(k)$, $T(k)$ and $\beta(k^2)$ terms in Eq.~(\ref{Eq_H_Vafek}). The effects of different terms are summarized in Tab.\ref{Tab_Wan}.
}}
\end{figure*}

Compared to the BM model Eq.~(\ref{Eq_H_BM}), 
in addition to the Dirac term $h_{\rm D}(k)$ and the local interlayer tunneling $T$ (which is $k$-independent), the fully relaxed continuum model introduces several other terms. To simplify, we label them based on their order of momentum dependence:
the momentum-independent in-plane strain field $A$, its first-order derivative terms $A(k)$ and second-order derivative terms $\beta(k^2)$, and the non-local interlayer tunnelling $T(k)$. Specifically,
\begin{equation}
\label{Eq_H_Vafek}
\begin{split}
H_{\rm intra} &= h_{\rm D}(k) + A + A(k) + \beta(k^2), \\
H_{\rm inter} &= T + T(k).
\end{split}
\end{equation}
Note that $T(k)$ in the fully relaxed model has a different form from the one in the generalized BM model in Eq.~(\ref{Eq_nonlocalT}).
By selectively deactivating these terms within the fully relaxed model, we provide a detailed comparison of the band structures and $\sigma_{\rm H}^{1}$, and gain intriguing insights on the effects of each term. 
Fig.~\ref{fig_SK} and Fig.~\ref{fig_Wan} provide a comprehensive view of these changes with Slater-Koster and Wannier model parameters respectively.
We summarize their combined impact on the band structure and $\sigma_{\rm H}^{1}$ in Tab.~\ref{Tab_SK} and Tab.~\ref{Tab_Wan}.
In the order of influence: the momentum-independent in-plane strain field $A$ greatly suppresses the flat band bandwidth to $\sim 10$ meV, changes flat band chirality, and most importantly flips the sign of $\sigma_{\rm H}^{1}$ on the hole-doping side, as well as increases the magnitude of $\sigma_{\rm H}^{1}$; $A(k)$ lowers the $\gamma$ point energies by around $1$ meV but compeletely reverses the effects of $A$ on $\sigma_{\rm H}^{1}$; the $T(k)$ term uplifts the $\gamma$ point energies by around $6-8$ meV.

\begin{table*}
\centering
\renewcommand{\arraystretch}{1.3} 
\begin{tabularx}{1\textwidth}{|c|c|X|}
\hline
\multicolumn{3}{|c|}{Slater-Koster tight-binding model parameters} \\
\hline
The term in $H$ & Figures to compare & Effects of the term \\
\hline
\multirow{3}{*}{$A$} & 
\multirow{3}{*}{\parbox{4.0cm}{Fig.~\ref{fig_SK}(a) vs. Fig.~\ref{fig_SK}(c) \\ Fig.~\ref{fig_SK}(b) vs. Fig.~\ref{fig_SK}(e)}} & 
\multirow{3}{=}{ 
\vspace{-4pt}
\begin{itemize}[left=5pt,topsep=0pt,partopsep=0pt,parsep=0pt,itemsep=0pt]
 \item Suppresses the flat band bandwidth from $\sim 80$ meV to $\sim 10$ meV.
 \item Changes the flat band chirality.
 \item Flips the sign of $\sigma_{\rm H}^{1}$ on the hole-doping side.
 \item Increases the magnitude of $\sigma_{\rm H}^{1}$ from $\sim 10 e^2/h$ to $\sim 60 e^2/h$.
\end{itemize}
} \\
& & \\
& & \\
& & \\
\hline
\multirow{4}{*}{$A(k)$} & 
\multirow{4}{*}{\parbox{4.0cm}{Fig.~\ref{fig_SK}(c) vs. Fig.~\ref{fig_SK}(d) \\
Fig.~\ref{fig_SK}(e) vs. Fig.~\ref{fig_SK}(f)}} & 
\multirow{4}{=}{ 
\vspace{0pt}
\begin{itemize}[left=5pt,topsep=0pt,partopsep=0pt,parsep=0pt,itemsep=0pt]
 \item Induces a small particle-hole asymmetry in the flat bands by lowering the $\gamma$ point energies by an energy scale $1-2$ meV.
 \item Completely reverses the effects of $A$ on $\sigma_{\rm H}^{1}$ described above.
\end{itemize}
} \\
& & \\
& & \\
& & \\
\hline
\multirow{3}{*}{$T(k)$} & 
\multirow{3}{*}{\parbox{4.0cm}{
Fig.~\ref{fig_SK}(a) vs. Fig.~\ref{fig_SK}(b) \\
Fig.~\ref{fig_SK}(c) vs. Fig.~\ref{fig_SK}(e) \\
Fig.~\ref{fig_SK}(d) vs. Fig.~\ref{fig_SK}(f)}} & 
\multirow{3}{=}{ 
\vspace{-0pt}
\begin{itemize}[left=5pt,topsep=0pt,partopsep=0pt,parsep=0pt,itemsep=0pt]
  \item Induces the particle-hole asymmetry in the flat bands by uplifting the $\gamma$ point energies by an energy scale of $\sim 8$ meV.
 \item The effect of $T(k)$ on $\sigma_{\rm H}^{1}$ is complex.
\end{itemize}
} \\
& & \\
& & \\
& & \\
\hline
\multirow{2}{*}{$\beta(k^2)$} & 
\multirow{2}{*}{\parbox{4.0cm}{
Fig.~\ref{fig_SK}(f) vs. Fig.~\ref{fig_relaxed}(a-b)}} & 
\multirow{2}{=}{ 
\vspace{-3pt}
\begin{itemize}[left=5pt,topsep=0pt,partopsep=0pt,parsep=0pt,itemsep=0pt]
\item Negligible effects on flat band spectrum, DOS and $\sigma_{\rm H}^{1}$.
\end{itemize}
} \\
& & \\
\hline
\end{tabularx}
\caption{Effects of different terms in Hamiltonian Eq.(\ref{Eq_H_Vafek}) on flat band spectrum and $\sigma_{\rm H}^{1}$, using Slater-Koster model parameters.}
\label{Tab_SK}
\end{table*}

\begin{table*}
\centering
\renewcommand{\arraystretch}{1.2} 
\begin{tabularx}{1.0\textwidth}{|c|c|X|}
\hline
\multicolumn{3}{|c|}{Wannier tight-binding model parameters} \\
\hline
The term in $H$ & Figures to compare & Effects of the term \\
\hline
\multirow{4}{*}{$A$} & 
\multirow{4}{*}{\parbox{4.0cm}{
Fig.~\ref{fig_Wan}(a) vs. Fig.~\ref{fig_Wan}(c) \\ Fig.~\ref{fig_Wan}(b) vs. Fig.~\ref{fig_Wan}(e)}} & 
\multirow{4}{=}{ 
\vspace{-8pt}
\begin{itemize}[left=5pt,topsep=0pt,partopsep=0pt,parsep=0pt,itemsep=0pt]
 \item Suppresses the flat band bandwidth from $\sim 30$ meV to $\sim 15$ meV.
 \item Changes the flat band chirality.
 \item Flips the sign of $\sigma_{\rm H}^{1}$ on the hole-doping side.
 \item Slightly increases the magnitude of $\sigma_{\rm H}^{1}$ from $\sim 10 e^2/h$ to $\sim 25 e^2/h$.
\end{itemize}
} \\
& & \\
& & \\
& & \\
\hline
\multirow{4}{*}{$A(k)$} & 
\multirow{4}{*}{\parbox{4.0cm}{Fig.~\ref{fig_Wan}(c) vs. Fig.~\ref{fig_Wan}(d) \\
Fig.~\ref{fig_Wan}(e) vs. Fig.~\ref{fig_Wan}(f)}} & 
\multirow{4}{=}{ 
\vspace{-8pt}
\begin{itemize}[left=5pt,topsep=0pt,partopsep=0pt,parsep=0pt,itemsep=0pt]
 \item Lowers the $\gamma$ point energies by an energy scale $< 1$ meV.
 \item Completely reverses the effects of $A$ on $\sigma_{\rm H}^{1}$ described above.
\end{itemize}
} \\
& & \\
& & \\
& & \\
\hline
\multirow{4}{*}{$T(k)$} & 
\multirow{4}{*}{\parbox{4.0cm}{
Fig.~\ref{fig_Wan}(a) vs. Fig.~\ref{fig_Wan}(b) \\
Fig.~\ref{fig_Wan}(c) vs. Fig.~\ref{fig_Wan}(e) \\
Fig.~\ref{fig_Wan}(d) vs. Fig.~\ref{fig_Wan}(f)}} & 
\multirow{4}{=}{ 
\vspace{-8pt}
\begin{itemize}[left=5pt,topsep=0pt,partopsep=0pt,parsep=0pt,itemsep=0pt]
 \item Induces the particle-hole asymmetry in the flat bands by uplifting the $\gamma$ point energies by an energy scale of $\sim 6$ meV.
 \item The effect of $T(k)$ on $\sigma_{\rm H}^{1}$ is complex.
\end{itemize}
} \\
& & \\
& & \\
& & \\
\hline
\multirow{2}{*}{$\beta(k^2)$} & 
\multirow{2}{*}{\parbox{4.0cm}{
Fig.~\ref{fig_Wan}(f) vs. Fig.~\ref{fig_relaxed}(e-f)}} & 
\multirow{2}{=}{ 
\vspace{-3pt}
\begin{itemize}[left=5pt,topsep=0pt,partopsep=0pt,parsep=0pt,itemsep=0pt]
\item Negligible effects on flat band spectrum, DOS and $\sigma_{\rm H}^{1}$.
\end{itemize}
} \\
& & \\
\hline
\end{tabularx}
\caption{Effects of different terms in Hamiltonian Eq.(\ref{Eq_H_Vafek}) on flat band spectrum and $\sigma_{\rm H}^{1}$, using Wannier model parameters.}
\label{Tab_Wan}
\end{table*}

\end{document}